\documentclass[preprint,eqsecnum,aps,prd,nofootinbib]{revtex4}
\usepackage{graphicx}

\usepackage{amsmath}
\usepackage{enumerate}

\def\ben{\begin{equation}}
\def\een{\end{equation}}
\def\bena{\begin{eqnarray}}
\def\eena{\end{eqnarray}}

\renewcommand{\H}{{\mathcal H}}

\newcommand{\I}{{\mathcal I}}
\newcommand{\K}{{K}}

\newcommand{\tg}{{\tilde g}}
\newcommand{\tnabla}{{\tilde \nabla}}
\newcommand{\tn}{{\tilde n}}
\newcommand{\tgam}{{\tilde \gamma}}

\newcommand{\R}{{\mathcal R}}

\begin{document}
\title {Energy Bounds in Designer Gravity}

\author{Aaron J. Amsel\footnote{\tt amsel@physics.ucsb.edu} and Donald Marolf\footnote{\tt marolf@physics.ucsb.edu}}

\affiliation{Physics Department, UCSB, Santa Barbara, CA 93106, USA}

\begin{abstract}
We consider asymptotically anti-de Sitter gravity coupled to tachyonic scalar fields with mass at or slightly
above the Breitenlohner-Freedman bound in $d \ge 4$ spacetime dimensions.  The boundary conditions in these ``designer gravity'' theories are defined in terms
of an arbitrary function $W$.  We give a general argument that the Hamiltonian generators of asymptotic symmetries for such systems will be finite,
and proceed to construct these generators using the covariant phase space method.  The direct calculation confirms that the generators
are finite and shows that they take the form of the pure gravity result plus additional contributions from the scalar fields.
By comparing the generators to the spinor charge, we derive a lower bound on the gravitational energy when i) $W$ has a global minimum,
ii) the Breitenlohner-Freedman bound is not saturated, and iii) the scalar potential $V$ admits a certain type of ``superpotential."

\end{abstract}

\maketitle

\section{Introduction}
The AdS/CFT correspondence \cite{M} has provoked much recent interest in spacetimes with anti-de Sitter boundary conditions.  The bulk side of such a duality is described by gravity coupled to various matter fields, typically including tachyonic scalars \cite{AdS5, AdS5krn, AdS7}.   When these scalars have
mass at or slightly above the Breitenlohner-Freedman (BF) bound, the analysis of \cite{BF} suggests that the energy is bounded below under
a range of boundary conditions on the scalar field.
It was further found in \cite{iw} that requiring the wave equation to have a sufficiently well-posed deterministic evolution law under general boundary conditions
restricts the scalar field mass to be in this same range.

Of course, the details of the theory will depend on the particular
boundary condition chosen. It is convenient to specify the
boundary condition through the choice of an arbitrary function
$W$, which under the AdS/CFT duality is expected to correspond to
a certain potential term in the Lagrangian of the dual CFT
\cite{witten, bss, ss}. On the bulk side, the choice of $W$
determines the existence and masses of certain solitons
\cite{thgh, sol1, hh}.  As a result, such models were termed
``designer gravity'' theories in  \cite{thgh}.  Cosmic censorship
\cite{cosmic},  black hole hair \cite{hair}, and  other
interesting  issues  \cite{three, other} have also been
investigated within designer gravity.

In this paper, we will be concerned with the construction of
conserved charges in asymptotically anti de-Sitter spacetimes
under the general scalar boundary conditions mentioned above.  We
consider the case of $d \ge 4$ spacetime dimensions.  Our results
overlap with the recent work  \cite{hmtz} which constructs these
generators using the Regge-Teitelboim method \cite{rt}.  Earlier,
such generators were constructed in \cite{three} for the case
$d=3$ (and a particular mass of the scalar field), and in
\cite{BFS1,BFS2,saturate} for the $d \ge 4$ case where the BF
bound is saturated.

We will not follow the Regge-Teitelboim approach here.
Instead, we adopt the
covariant phase space method \cite{WL,wi,wz}, and our focus will be on positivity properties of the resulting energy.  Since both our approach
and that of \cite{hmtz} lead to generators of time translations, they must agree up to an additive constant.  The same argument also applies
(see \cite{HIM2}) to any energy constructed via the boundary stress tensor methods of, e.g., \cite{skenderis,bk}.

Positivity properties of the energy have implications for the
stability of the theory. In particular, when $W$ is bounded below,
the dual CFT is expected to be stable.  This suggests that the
same should be true on the gravity side. A positive energy theorem
with trivial scalar boundary conditions was proven in \cite{t}
(see also \cite{ST,FNSS} for extensions). The case of ${\cal N} =
8$, $d =4$ gauged supergravity with scalar fields of mass $m^2 =
-2$ was analyzed in \cite{hh}. The goal of this paper is to
generalize the results of \cite{hh} to any spacetime dimension $d
\geq 4$ and to any mass in the range $m^2_{BF} \le m^2 < m^2_{BF}
+1$.  To simplify the analysis, we consider only even scalar
potentials.

The plan of this paper is as follows.  In section~\ref{prelim}, we
describe our choice of asymptotic conditions for the metric and
the scalar field. Section~\ref{structure} contains a brief general
argument demonstrating that the symplectic structure is finite,
which then implies that the Hamiltonian generators in our theory
must be finite as well. In section~\ref{metric}, we analyze the
consequences of combining our asymptotic conditions with
Einstein's equation.  We derive explicit expressions for the
Hamiltonian generators of asymptotic symmetries in
section~\ref{charges}. In section~\ref{spinor}, we use the
positivity of the spinor charge to obtain a lower bound on the
gravitational energy for designer gravity theories such that i) $W$ has a global minimum,
ii) the Breitenlohner-Freedman bound is not saturated, and iii) the scalar potential $V$ admits a certain type of ``superpotential."
These results are summarized and discussed
in section~\ref{disc}.

\section{Preliminaries}
\label{prelim}

Our notations and conventions follow those of \cite{wald}.  The
dimension of spacetime is denoted $d$, and our analysis is
restricted to $d\geq 4$.  The signature of the metric is $(-++
\ldots )$.  Indices on tilde tensor fields $\tilde t_{abc\dots}$ are
raised and lowered with the unphysical metric $\tilde g_{ab} =
\Omega^2 g_{ab}$ and its inverse $\tilde g^{ab}$.  The Levi-Civita
tensor associated to $\tilde g_{ab}$ is denoted $\tilde
\epsilon_{a_1 a_2 \ldots a_d}$. The AdS radius and $8 \pi G$ are set
to one.

We consider a theory of gravity coupled to a tachyonic
($m^2 < 0$) scalar field with mass in the range
\begin{equation}
\label{range}
m^2_{BF} \le m^2 < m^2_{BF} +1,
\end{equation}
where $m^2_{BF} =-(d-1)^2/4$.
The Lagrangian density (written as a $d$-form) for such a theory with minimal coupling is
\begin{equation}
\label{theory}
{\bf L} = \frac{1}{2} \, d^d x \sqrt{ - g} \, [R - (\nabla \phi)^2 -
2V(\phi)] ,
\end{equation}
where the scalar potential is $V(\phi) = \Lambda +\bar{V}(\phi)$ and we assume
\begin{equation}
\bar{V}(\phi)
= \frac{1}{2} m^{2} \phi^{2} + \frac{1}{4} c_4 \phi^4 + O(\phi^6) + \ldots.
\end{equation}
The $O(\phi^6)+ \ldots$ terms in the scalar
potential are irrelevant to calculating the Hamiltonian generators
in the regime of interest~(\ref{range}). However, $O(\phi^4)$ terms
must be taken into account for a limited range of masses that
occurs only in $d = 4$. A similar statement would hold for $
O(\phi^3)$ and $O(\phi^5)$ terms, but for simplicity we consider
only even potentials here. (See \cite{hmtz} for a construction of
Hamiltonian generators that allows odd terms in the potential).

Our asymptotic conditions, following those of \cite{HIM}, are described below:
\begin{enumerate}
\item One can attach a boundary, $\I \cong R \times S^{d-2}$
to $M$ such that $\tilde M = M \cup \I$ is a manifold
with boundary.
\item
On $\tilde M$, there is a $(d-1)$-times continuously differentiable metric $\tg_{ab}$ and a smooth
function $\Omega$ such that $g_{ab} = \Omega^{-2} \tg_{ab}$, with $\Omega = 0$ and
\ben
\tn_a \equiv \tnabla_a \Omega \neq 0
\een
at points of $\I$. We also require that the metric $\tilde h_{ab}$ on $\I$ induced by $\tg_{ab}$
is the Einstein static universe,
\ben
\tilde h_{ab} \, dx^a dx^b = -dt^2 + d \sigma^2,
\een
where $d\sigma^2$ is the line element of the unit sphere $S^{d-2}$.
\end{enumerate}
An example of a spacetime satisfying these conditions is of course exact AdS space.  In global coordinates, the metric is
\begin{equation}
\label{pureads}
ds^{2}_{0} = - \left(1 + \frac{r^2}{\ell^2}\right)\, dt^2 + \frac{dr^2}{1 + r^2/\ell^2}
+ r^2 d\sigma^2 \, .
\end{equation}
We will set the AdS radius $\ell =1$ so that $\Lambda = -\frac{1}{2} (d-1)(d-2)$.  After performing the change of coordinates $r = \Omega^{-1}-\Omega /4$ for
$\Omega$ a smooth and positive function,
the unphysical metric $\tg_{ab} = \Omega^{2} g_{ab}$ can be put in the form
\begin{equation}
\label{ds0}
d\tilde s^{2}_{0} = d\Omega^2 - \left(1 + \frac{1}{4} \Omega^2 \right)^2 dt^2 +\left(1 - \frac{1}{4} \Omega^2 \right)^2 d\sigma^2 \, ,
\end{equation}
which has the properties required above.

To determine the asymptotic behavior of the
scalar field, we study the equation of motion
\begin{equation}
\label{waveeq}
\nabla^a \nabla_a \phi - \frac{dV}{d\phi} = 0 \,.
\end{equation}
For most values of $m^2$ and $d$, it is sufficient to solve the linear
equation in exact AdS space.
One finds for $m^2 \neq m^{2}_{BF}$ that
\begin{equation}
\label{phi}
\phi = \alpha \Omega^{\lambda_{-}} + \beta \Omega^{\lambda_{+}}  + \dots,
\end{equation}
where
\begin{equation}
\label{roots}
\lambda_{\pm} = {d-1 \pm \sqrt{(d-1)^2+ 4 m^2}\over 2} \,
\end{equation}
and the coefficients $\alpha, \beta$ are smooth functions on $\I$.  It will be useful to observe that eq.~(\ref{range}) implies
$d-3 < 2\lambda_{-} \leq d-1$ and $d-1 \leq 2\lambda_{+} < d+1$.

Corrections to eq.~(\ref{phi}) can arise from coupling to the metric
(i.e. back reaction) and/or from the $\phi^4$ term in the potential.
It turns out that the leading correction from both of these effects
is at $O(\Omega^{3\lambda_-})$, so this term is of the same or lower
order than the second term in eq.~(\ref{phi}) when $4\lambda_- \leq
d-1$, which only occurs in $d=4$. Hence, for these special cases we
must be more careful when solving the equation of
motion~(\ref{waveeq}).  We assume a series solution for $\phi$ to
solve the nonlinear wave equation and find for $4\lambda_- < d-1$
the result
\begin{equation}
\phi = \alpha \Omega^{\lambda_{-}} + \gamma_1 \alpha^3 \Omega^{3\lambda_{-}} + \beta \Omega^{\lambda_{+}}  +\ldots \,,
\end{equation}
where
\begin{equation}
\label{gam1}
\gamma_1 = \frac{c_g +c_4}{2\lambda_- (4\lambda_- +1 -d)} \,.
\end{equation}
Here $c_g$ is a constant that arises from gravitational back reaction (see section \ref{metric}) and is given by
\begin{equation}
c_g = \frac{(d-1)\lambda_-^2}{2(d-2)} \,.
\end{equation}
For $d = 4$ and $4\lambda_- = 3$ we proceed similarly, but  now logarithmic terms arise in the series solution for $\phi$.  The result is
\begin{equation}
\phi = \alpha \Omega^{\lambda_{-}} + \gamma_2 \alpha^3 \Omega^{3\lambda_{-}} \, \textrm{log} \, \Omega + \beta \Omega^{\lambda_{+}} + \ldots \, ,
\end{equation}
where
\begin{equation}
\label{gam2}
\gamma_2 = \frac{2}{3}(c_g+c_4)\,.
\end{equation}

For $m^2 = m^{2}_{BF}$, the roots~(\ref{roots}) are degenerate and the solution becomes
\begin{equation}
\label{phibf}
\phi = -\hat{\alpha} \Omega^{\lambda}\,\textrm{log}\,\Omega + \hat{\beta} \Omega^{\lambda} + \dots \,,
\end{equation}
where $\lambda = (d-1)/2$.

A valid boundary condition must ensure that no symplectic flux flows through the boundary $\I$.  In particular, this permits
the Hamiltonian generators to be well defined \cite{wz}.  As we will see in section~\ref{structure} below, for each case above it is sufficient to fix
a functional relationship  $\beta = \beta(\alpha)$.  We parametrize this relationship via a smooth function $W(\alpha)$ satisfying
\begin{equation}
\label{dW}
\beta \equiv \frac{dW}{d\alpha} \,.
\end{equation}

\section{The Symplectic Structure}
\label{structure}
Before going through the detailed calculation of the Hamiltonian generators, we first give a concise argument that they must be finite. Our argument
uses the symplectic structure of the theory, which we now define.  We will also verify below that the symplectic flux through the boundary vanishes.

In general, the variation of the Lagrangian density can be written in the form
\begin{equation}
\label{dL}
\delta {\bf L} = {\bf F} \cdot \delta \Phi + d\boldsymbol{\theta} \, ,
\end{equation}
where $\Phi = (g_{ab}, \phi)$.  The equations of motion of the theory are ${\bf F} = 0$, and $d\boldsymbol{\theta}$ corresponds to the boundary term
that would arise from integrating by parts.  $\boldsymbol{\theta}$ is referred to as the symplectic potential, and for our theory~(\ref{theory}), it follows that
\begin{equation}
\label{theta}
\theta_{a_1 \dots a_{d-1}} =  \left[ \frac{1}{2} (
\nabla^c \delta g_c{}^b - \nabla^b \delta g_c{}^c) -
\delta \phi \nabla^b \phi \right] \epsilon_{ba_1 \dots a_{d-1}} \,.
\end{equation}
The symplectic current is given as the antisymmetrized variation of
the symplectic potential,
\begin{equation}
\label{current} \boldsymbol{\omega}(\Phi; \delta_1 \Phi, \delta_2
\Phi) = \delta_1 \boldsymbol{\theta}(\Phi; \delta_2 \Phi) - \delta_2
\boldsymbol{\theta}(\Phi; \delta_1 \Phi) \, ,
\end{equation}
and the symplectic structure is then defined as
\begin{equation}
\label{sform}
\sigma_{\Sigma}(\Phi; \delta_1 \Phi, \delta_2 \Phi) =
\int_{\Sigma} \boldsymbol{\omega}(\Phi; \delta_1 \Phi, \delta_2
\Phi) \, ,
\end{equation}
where $\Sigma$ is a surface whose boundary $S_\infty$ is a cut of
$\I$.  It follows from definition~(\ref{current}), eq.~(\ref{dL}),
and the equality of mixed partial derivatives that
$d\boldsymbol{\omega} = 0$ when the linearized equations of motion hold.  Stokes's theorem can then be used to show that the
symplectic structure is conserved, i.e. it is independent of the choice of
surface $\Sigma$, when the pull-back of $\boldsymbol{\omega}$ to
$\I$ vanishes \cite{WL, wi}.

It is convenient to write the symplectic current as
$\boldsymbol{\omega} = \boldsymbol{\omega}^{\phi} + \boldsymbol{\omega}^{G}$.
Here $\boldsymbol{\omega}^{\phi}$ arises
from terms depending on the explicit variation of the scalar field.  Eqs.~(\ref{current}), (\ref{theta}), and (\ref{phi}) imply
that to leading order in $\Omega$ one has
\begin{equation}
\label{wphi}
\omega^{\phi}_{a_1 \dots a_{d-1}} = \left(\delta_1 \alpha
\tnabla^b (\delta_2\alpha)- \delta_2 \alpha
\tnabla^b (\delta_1\alpha)\right) \Omega^{2\lambda_- +2
-d}\tilde{\epsilon}_{ba_1
\dots a_{d-1}} +... \, ,
\end{equation}
where we have assumed that the variations commute, $\delta_1
\delta_2 - \delta_2 \delta_1 = 0$.  The requirement for the integral over $\Omega$ in eq.~(\ref{sform}) to not diverge at $\Omega \to 0$ is then
$2\lambda_- > d-3$ or equivalently $m^2 < m^{2}_{BF} +1$, which agrees with the range~(\ref{range}).
The terms contained in $\boldsymbol{\omega}^G$ are due to the variation of the metric and are given by
\cite{wz}
\begin{equation}
\label{wdef}
\omega^{G}_{a_1 \dots a_{d-1}} = \frac{1}{2}
P^{abcdef}(\delta_2 g_{bc} \nabla_d \delta_1 g_{ef} - \delta_1
g_{bc} \nabla_d \delta_2 g_{ef}) \epsilon_{a a_1 \dots a_{d-1}} \, ,
\end{equation}
where
\begin{equation}
P^{abcdef}
  = g^{ae} g^{fb} g^{cd}
  - \frac{1}{2} g^{ad} g^{be} g^{fc}
  - \frac{1}{2}g^{ab} g^{cd} g^{ef}
  - \frac{1}{2}g^{bc} g^{ae} g^{fd}
  + \frac{1}{2}g^{bc} g^{ad} g^{ef}.
\end{equation}
In the next section, we will show by expanding the metric in powers of $\Omega$ that Einstein's equation implies
\begin{equation}
\label{varg}
\delta \tg_{ab} = - \frac{1}{2(d-2)} \Omega^{2\lambda_-} \delta(\alpha^2)  (\tilde h_{ab})_0 + \ldots \,,
\end{equation}
where $(\tilde h_{ab})_0$ is the metric of the Einstein static universe.  It is then
straightforward to rewrite eq.~(\ref{wdef}) in terms of the unphysical metric and substitute the variation~(\ref{varg}) to obtain
$\boldsymbol{\omega}^{G} \sim O(\Omega^{4\lambda_- + 2 -d})$.  Once again, the integral over $\Omega$ of this term in the symplectic structure
converges as $\Omega \to 0$.

Now, let $\xi^a$ be a vector field representing an asymptotic symmetry,
in the sense that $\xi^a$ is asymptotically a Killing vector field and that the corresponding diffeomorphism is a symmetry of the
covariant phase space.
Any Hamiltonian generator of an asymptotic symmetry must satisfy
\begin{equation}
\label{gendef}
\delta \H_\xi =
\sigma_{\Sigma}(\Phi; \delta \Phi, {\pounds_\xi} \Phi) \, .
\end{equation}
This is essentially Hamilton's equation of motion for ``time''
translation generated by $\xi$.  The charge associated to $\xi$ is conserved, since the symplectic structure is independent
of $\Sigma$.
The above arguments then show
that variations of the Hamiltonian generators for our theory are finite everywhere in the range~(\ref{range}).  Thus, any formal expression
whose variation satisfies~(\ref{gendef}) must be finite, up to an arbitrary additive constant.  Subtracting this constant leaves a finite
generator $\H_\xi$.

As argued in~\cite{wz}, the generator $\H_\xi$ cannot exist unless the consistency condition $\xi \cdot \boldsymbol{\omega} \mid_{\I} \, = 0$ is satisfied.
Since $\tn^a \tnabla_a \alpha =0 $, we find using the above definitions that
\begin{equation}
\label{wphi2}
\omega^{\phi}_{a_1 \dots a_{d-1}} \mid_\I \,= (\lambda_+-\lambda_-)\left(\delta_1 \beta
\delta_2\alpha- \delta_2 \beta
\delta_1\alpha \right) \tn^b
\tilde{\epsilon}_{ba_1\dots a_{d-1}} \, .
\end{equation}
This vanishes because the boundary condition~(\ref{dW})
implies $\delta \beta = W''(\alpha) \delta \alpha$.
Substituting the metric variation~(\ref{varg}) into eq.~(\ref{wdef}) similarly shows that $\boldsymbol{\omega}^G \mid_\I \,= 0$.
So the consistency condition is indeed satisfied.

\section{Metric Asymptotics}
\label{metric}

We will now calculate back reaction effects on the metric due to the presence of the scalar field.  We proceed by combining
the asymptotic conditions described above with Einstein's equation.  In the tradition of \cite{FG}, we expand the unphysical metric
in powers of $\Omega$.  The calculation is organized as in \cite{HIM}.

Defining the tensor
\begin{equation}
\label{Sdef}
\tilde S_{ab} = \frac{2}{d-2} \tilde R_{ab} - \frac{1}{(d-1)(d-2)} \tilde R \tg_{ab} \,,
\end{equation}
Einstein's equation may be written as
\begin{equation}
\label{einstein}
\tilde S_{ab} = - 2 \Omega^{-1} \tilde \nabla_a \tn_b + \tilde{L}_{ab} \,,
\end{equation}
where
\begin{equation}
\label{Labdef}
\tilde L_{ab} = \frac{2}{d-2}\left[ T_{ab} - \frac{1}{d-1} g_{ab} T \right] \, ,
\end{equation}
and the matter stress energy tensor is
\begin{equation}
\label{phiT}
T_{ab} = \nabla_a \phi \nabla_b \phi - g_{ab} \left[\frac{1}{2} \nabla^c \phi \nabla_c \phi
+ \bar{V}(\phi)\right].
\end{equation}
The conformal factor $\Omega$ is chosen such that $\tn_a \equiv \tnabla_a \Omega$ is spacelike, unit, and normal to
$\I$.  It is then convenient to write the unphysical metric in the Gaussian normal form
\begin{equation}
\tg_{ab} = \tn_a \tn_b + \tilde{h}_{ab} \,.
\end{equation}
Einstein's equation~(\ref{einstein}) can be split into its components parallel and normal to surfaces of constant $\Omega$.
We obtain the constraint equations
\begin{eqnarray}
-\tilde \R - \tilde \K_{ab} \tilde \K^{ab} + \tilde \K^2 + 2(d-2) \Omega^{-1} \tilde \K
&=& \tilde{L}_{ab} \tilde{h}^{ab}  \\
\tilde D^a \tilde \K_{ab} - \tilde D_b \tilde \K &=&
\tilde{L}_{cd} \tilde{n}^{c} \tilde{h}^d{}_{b} \,,
\end{eqnarray}
where $\tilde D_a$ is the derivative operator associated with $\tilde h_{ab}$,
$\tilde \K_{ab} \equiv - \tilde h_a{}^c \tilde h_b{}^d \tilde \nabla_c \tn_d = -\tilde\nabla_a \tn_b$
is the extrinsic curvature, and $\tilde \R_a{}^b$ is the intrinsic Ricci tensor.
The evolution equations are
\begin{eqnarray}
\label{evol1}
\frac{d}{d\Omega} \tilde \K_a{}^b &=&
-\tilde \R_a{}^b + \tilde \K \tilde \K_a{}^b + \Omega^{-1}(d-2) \tilde
\K_a{}^b + \Omega^{-1} \tilde \K \delta_a{}^b \nonumber\\
&&+ \frac{1}{2} (d-2) \tilde{h}_{a}{}^{c} \tilde{h}^{bd} \tilde{L}_{cd} +
\frac{1}{2} \tilde{L}_{cd} \tilde{g}^{cd} \delta_a{}^b \\
\label{evol2}
\frac{d}{d\Omega} \tilde h_{ab} &=& -2 \tilde h_{bc} \tilde \K_a{}^c.
\end{eqnarray}
In these equations, differentiation with respect to $\Omega$ should be interpreted as the Lie derivative $\pounds_\tn$.  To solve eqs.~(\ref{evol1}) and (\ref{evol2}), we write
\begin{equation}
\label{expand}
\tilde h_{ab} = \sum_{j}\Omega^{\lambda_j} (\tilde h_{ab})_{\lambda_j}, \quad
\tilde K = \sum_{j} \Omega^{\lambda_j} (\tilde K)_{\lambda_j}, \ldots \textrm{etc.}
\end{equation}
Here the coefficients $(\tilde h_{ab})_{\lambda_j}$, $(\tilde K)_{\lambda_j}$,...etc. are independent of $\Omega$,
and the labels $\lambda_j$ are meant to indicate that powers in the expansions will not in general be integers.
In particular, the powers will take the form $(n\lambda_- + m)$ for integers $n,m$.
To determine the $(\tilde{h}_{ab})_j$, we insert the expansions~(\ref{expand}) into the evolution equations
and solve for the coefficients order by order in $\Omega$.
The result is the following series of recursion relations:
\begin{eqnarray}
(d-2-\lambda_j)(\tilde p_a{}^b)_{\lambda_j} &=&
(\tilde \R_a{}^b)_{\lambda_j-1} - \frac{1}{d-1}(\tilde \R)_{\lambda_j-1} \delta_a{}^b
\nonumber \\
&&- \sum_{m} (\tilde \K)_{\lambda_m} (\tilde p_a{}^b)_{\lambda_j-1-\lambda_m} -\frac{d-2}{2} (\tilde \tau_a{}^b)_{\lambda_j-1}
\label{recursionp}
\\
(2d-3-\lambda_j)(\tilde \K)_{\lambda_j} &=& (\tilde \R)_{\lambda_j-1}
- \sum_{m} (\tilde \K)_{\lambda_m} (\tilde \K)_{\lambda_j-1-\lambda_m}
\nonumber \\
&&-\frac{d-1}{2} (\tilde L_{cd})_{\lambda_j-1} \tn^c \tn^d - \frac{2d-3}{2}  (\tilde{L}_{cd} \tilde{h}^{cd})_{\lambda_j-1}
\label{recursionk}
\end{eqnarray}
\begin{equation}
\lambda_j(\tilde h_{ab})_{\lambda_j}
= -2\sum_{m}
  \Bigg[ (\tilde h_{bc})_{\lambda_m} (\tilde p_a{}^c)_{\lambda_j-1-\lambda_m}
         + \frac{1}{d-1} (\tilde h_{ab})_{\lambda_m} (\tilde \K)_{\lambda_j-1-\lambda_m} \Bigg].
\label{recursionh}
\end{equation}
Here we have defined $\tilde p_a{}^b$ as the traceless part of $\tilde K_a{}^b$ and $\tilde \tau_a{}^b$ as the traceless part of
$\tilde{h}_a{}^c \tilde{h}^{bd} \tilde{L}_{cd}$.  The initial conditions are
$(\tilde p_a{}^b)_0 = (\tilde \K)_0 = 0$ and $(\tilde{h}_{ab})_{0} dx^a dx^b = -dt^2 +d\sigma^2$.

Let us first consider the case $4\lambda_- > d-1$ and $m^2 > m^2_{BF}$. 
As noted in section \ref{prelim}, the first bound follows immediately from eqs.~(\ref{range}) and  (\ref{roots}) for $d \geq5$.
Inserting the asymptotic
expansion of $\phi$ (\ref{phi}) into the definition of $\tilde{L}_{ab}$ leads to
\begin{eqnarray}
\label{Lab}
\tilde L_{ab} &=& \frac{2 \lambda_{-} \alpha^{2}}{d-2} \, \Omega^{2 \lambda_{-} -2} \left[\lambda_{-} \tn_{a} \tn_{b} - \frac{1}{2} \tilde{g}_{ab}\right]
+ \frac{4 \lambda_{-} \alpha}{d-2} \, \Omega^{2 \lambda_{-} -1} \, \tn_{( a} \tilde{\nabla}_{b)} \alpha \nonumber\\
&& +\frac{4m^{2} \alpha \beta}{d-2} \, \Omega^{d-3} \left[ \frac{1}{d-1} \tilde{g}_{ab}
- \tn_{a} \tn_{b}\right]+ \ldots \,.
\end{eqnarray}
Eqs.~(\ref{recursionk}) and (\ref{Lab}) indicate that the lowest order at which the nonzero stress energy tensor affects $\tilde K$
is $(\tilde{K})_{2\lambda_- - 1}$.  The $\tilde \R$ and $\tilde{K}^2$ terms on the right hand side of~(\ref{recursionk})
must then take the same value as in the $T_{ab} = 0$ case,
while the matter fields give a contribution
\begin{displaymath}
\frac{(d-1)\lambda_-}{2(d-2)} \, \alpha^2 \,.
\end{displaymath}
Proceeding similarly with eq.~(\ref{recursionp}), we find that the matter contribution to $(\tilde p_a{}^b)_{2\lambda_- -1}$ vanishes.
Then eq.~(\ref{recursionh}) implies that the resulting term in the unphysical metric is
\begin{displaymath}
-\frac{\alpha^2}{2(d-2)} \, \Omega^{2\lambda_-} (\tilde{h}_{ab})_0 \,,
\end{displaymath}
which must be added to eq.~(\ref{ds0}).
The $O(\Omega^{2 \lambda_{-} -1})$ term in $\tilde{L}_{ab}$ gives no contribution to $\tilde{h}_{ab}$ since $\tn^a \nabla_a  \alpha = 0$ and
$\tn^a \tilde{h}_{ab} =0$.
The recursion relations fail to determine $(\tilde{h}_{ab})_{d-1}$, but this coefficient can be written in terms of the electric part
of the Weyl tensor, defined  as
\begin{equation}
\tilde{E}_{ab} \equiv \frac{1}{d-3} \Omega^{3-d} \tilde C_{acbd} \tn^c \tn^d \, .
\end{equation}
The relation
\begin{equation}
\tilde C_{acbd} \tn^c \tn^d=
\frac{d}{d \Omega} \tilde K_{ab} - \Omega^{-1} \tilde K_{ab}
+ \tilde K_{ac} \tilde K^c{}_b - \frac{1}{2} \tilde h_a{}^c \tilde h_b{}^d \tilde L_{cd}
- \frac{1}{2} \tilde h_{ab} \tilde L_{cd} \tn^c \tn^d \,
\end{equation}
combined with eq.~(\ref{evol2}) results in
\begin{eqnarray}
\label{habd}
(\tilde{h}_{ab})_{d-1} = &-&\frac{2}{d-1} (\tilde{E}_{ab})_0 +\frac{2}{(d-1)(d-3)} (\tilde{K}_{ac} \tilde{K}^c{}_b)_{d-3} \cr &-& \frac{1}{(d-1)(d-3)}
(\tilde h_a{}^c \tilde h_b{}^d \tilde L_{cd} + \tilde h_{ab} \tilde L_{cd} \tn^c \tn^d)_{d-3} \,.
\end{eqnarray}
One can check that, so long as $4\lambda_- \neq d-1$ (a case treated separately below), the second term in eq.~(\ref{habd}) takes the same value
as in the case with $T_{ab} = 0$.
The third term may be directly worked out from eq.~(\ref{Lab}).
Then, the expression for the metric near infinity is
\begin{eqnarray}
\label{ds}
d\tilde s^2 &=& d\Omega^2 - \left[\left(1 + \frac{1}{4} \Omega^2 \right)^2 - \frac{\alpha^2}{2(d-2)}
\Omega^{2 \lambda_{-}} + \frac{4 m^{2} \alpha \beta}{(d-2)(d-1)^2}
\Omega^{d-1} \right] dt^2 \nonumber \\
&&+  \left[ \left(1 - \frac{1}{4} \Omega^2 \right)^2 -
\frac{\alpha^2}{2(d-2)} \Omega^{2 \lambda_{-}} + \frac{4 m^{2}
\alpha \beta}{(d-2)(d-1)^2} \Omega^{d-1}
\right]d\sigma^2 \nonumber \\
&&- \frac{2}{d-1}\Omega^{d-1} \tilde{E}_{ab} \, dx^a dx^b + \dots \,.
\end{eqnarray}
In general, there are corrections to eq.~(\ref{ds}) at $O(\Omega^{4 \lambda_{-}})$ that are
proportional to $\alpha^4$.  For the special cases $4 \lambda_{-} \leq d-1$, such $O(\Omega^{4 \lambda_{-}})$ terms are of the same or lower order
than the $\Omega^{d-1}$ terms, so we must keep track of them explicitly.  For $4 \lambda_{-} < d-1$ we find (setting $d=4$)
\begin{eqnarray}
\label{ds2}
d\tilde s^2 &=& d\Omega^2 - \left[\left(1 + \frac{1}{4} \Omega^2 \right)^2 - \frac{\alpha^2}{4}
\Omega^{2 \lambda_{-}} + A\alpha^4\Omega^{4\lambda_-} + \frac{2 m^{2} \alpha \beta}{9}
\Omega^{3} \right] dt^2 \nonumber \\
&&+  \left[ \left(1 - \frac{1}{4} \Omega^2 \right)^2 -
\frac{\alpha^2}{4} \Omega^{2 \lambda_{-}} + A\alpha^4\Omega^{4\lambda_-}+ \frac{2 m^{2}
\alpha \beta}{9} \Omega^{3}
\right]d\sigma^2 \nonumber \\
&&- \frac{2}{3}\Omega^{3} \tilde{E}_{ab} \, dx^a dx^b + \dots \, ,
\end{eqnarray}
where
\begin{equation}
A = \frac{1-12\gamma_1}{32} \,.
\end{equation}
Only two cases remain.  One case is $4 \lambda_{-} = d-1 =3$, for which we find
\begin{eqnarray}
\label{ds3}
d\tilde s^2 &=& d\Omega^2  \cr
&-& \left[\left(1 +  \Omega^2/4  \right)^2 - \frac{\alpha^2}{4}
\Omega^{3/2} - \frac{3\gamma_2}{8} \alpha^4 \Omega^{3} \textrm{log} \, \Omega
- \left(\frac{3\alpha \beta}{8} -\left(\frac{\gamma_2}{6}-\frac{c_4}{18} +\frac{1}{128}\right)\alpha^4\right)
\Omega^{3} \right] dt^2 \nonumber \\
&+&  \left[ \left(1 - \Omega^2/4 \right)^2 -
\frac{\alpha^2}{4} \Omega^{3/2} - \frac{3\gamma_2}{8} \alpha^4 \Omega^{3} \textrm{log} \, \Omega
- \left(\frac{3\alpha \beta}{8} -\left(\frac{\gamma_2}{6}-\frac{c_4}{18} +\frac{1}{128}\right)\alpha^4\right)
\Omega^{3}
\right]d\sigma^2 \nonumber \\
&-& \frac{2}{3}\Omega^{3} \tilde{E}_{ab} \, dx^a dx^b + \dots \,.
\end{eqnarray}
The final case occurs when the BF bound is saturated. The asymptotic behavior of the metric in this instance is given in eq.~(\ref{dsBF}) of the appendix.

\section{Conserved Charges}
\label{charges}

We now derive expressions for the Hamiltonian generators of asymptotic symmetries
using the covariant phase space formalism and the asymptotic behavior of the fields found in previous sections.  The calculation is best done by
first rewriting $\delta \H_\xi$ as a pure surface integral.

To this end, consider the Noether current defined by
\begin{equation}
{\bf J}_\xi = \boldsymbol{\theta} - \xi \cdot {\bf L} \,,
\end{equation}
where the centered dot indicates that $\xi^a$ is contracted with the
first index of $\bf L$.  When the equations of motion hold, $d{\bf
J}_\xi = 0$, so locally, ${\bf J}_\xi = d{\bf Q}_\xi$.  Here ${\bf
Q}_\xi$ is referred to as the Noether charge, and for our theory we
find
\begin{equation}
(Q_{\xi})_{a_1 \dots a_{d-2}} = -\frac{1}{2} \nabla^b \xi^c
\epsilon_{bca_1 \dots a_{d-2}} \, .
\end{equation}
By considering the variation $\delta{\bf J}_\xi$, the definitions in section \ref{structure} show that eq.~(\ref{gendef}) can be rewritten
as an integral over the boundary of $\Sigma$ \cite{wi},
\begin{equation}
\delta \H_\xi  = \int_{S_\infty} [\delta {\bf Q}_\xi - \xi \cdot \boldsymbol{\theta}] \,.
\end{equation}
Next we wish to calculate explicitly the quantity $\delta {\bf Q}_\xi - \xi \cdot  \boldsymbol{\theta}$.
We will find that each term diverges when considered separately, but that $\delta \H_\xi$ is indeed finite as predicted by the argument of section~\ref{structure}.
Given the result~(\ref{ds}), we assume variations of the form
\begin{eqnarray}
\label{phivar}
\delta \phi &=& \delta \alpha \Omega^{\lambda_{-}} + \delta \beta \Omega^{\lambda_{+}}  + \dots \\
\label{gvar}
\delta \tilde{g}_{ab} &=& - \frac{1}{2(d-2)} \Omega^{2\lambda_-} \delta(\alpha^2) (\tilde h_{ab})_0
+\frac{4 m^2}{(d-2)(d-1)^2} \Omega^{d-1} \delta (\alpha \beta) (\tilde h_{ab})_0
\nonumber\\ &&-\frac{2}{d-1} \Omega^{d-1} \delta \tilde{E}_{ab}+ \ldots \, .
\end{eqnarray}
Here we hold $\Omega$ fixed under the variation and, as usual, we defer the special cases~(\ref{ds2}) and (\ref{ds3}) for later treatment.
The Noether charge can be rewritten in terms of unphysical quantities as
\begin{equation}
(Q_\xi)_{a_1 \dots a_{d-2}}
= \Omega^{1-d} \tilde \epsilon_{bca_1 \dots a_{d-2}}
\tn^b \xi^c
-\frac{1}{2} \Omega^{2-d}
\tilde \epsilon_{bc a_1 \dots a_{d-2}}
\tilde \nabla^b \xi^c \, .
\end{equation}
Then, using
\begin{eqnarray}
\delta \tilde{\bf \epsilon} &=& \frac{1}{2} \tilde{g}^{ab} \delta \tilde{g}_{ab} \tilde{\bf \epsilon} \, ,
 \\ \delta \tn^a &=& -\tilde{g}^{ab} \tn^c \delta \tilde{g}_{bc} \, , \quad \textrm{and}
\\ \delta (\tilde \nabla^b \xi^c ) &=& -\tilde{g}^{bd} \tilde \nabla^f \xi^c \delta \tilde{g}_{df} + \tilde{g}^{be} \delta \tilde{\Gamma}^{c}_{ed} \xi^d \, ,
\end{eqnarray}
we obtain
\begin{equation}
(\delta Q_\xi)_{a_1 \dots a_{d-2}} \mid_{\I} = \left[\frac{2\lambda_{-}+1-d}{4(d-2)} \delta(\alpha^2) \Omega^{2\lambda_{-}-d+1} \tn^b \xi^c +
\tn^b \delta \tilde{E}^{c}{}_{d} \xi^d\right] \tilde \epsilon_{bca_1 \dots a_{d-2}} \,.
\end{equation}
Expanding (\ref{theta}) in powers of $\Omega$ yields
\begin{equation}
\theta_{a_1 \dots a_{d-1}} \mid_{\I} = \left[\frac{2\lambda_{-}+1-d}{4(d-2)} \delta(\alpha^2) \Omega^{2\lambda_{-}-d+1}  +
(\lambda_+-\lambda_-)\left(\frac{\lambda_-}{d-1} \delta(\alpha \beta)-\beta \delta \alpha \right) \right]\tn^b \tilde \epsilon_{ba_1 \dots a_{d-1}} \,.
\end{equation}
Note that both $\theta$ and $\delta Q_\xi$ contain terms diverging as $O(\Omega^{2\lambda_{-}-d+1})$ in the limit $\Omega \to 0$,
but that these terms cancel when we calculate $\delta \H_\xi$.
Let us choose $t^a = (\partial/\partial t)^a$ to be the unit timelike normal to $S_\infty$.  Writing the volume forms as
\begin{equation}
\tilde \epsilon_{bca_1 \dots a_{d-2}} \tn^b v^c= {}^{(d-1)} \tilde \epsilon_{ca_1 \dots a_{d-2}} v^c  = {}^{(d-2)}\tilde \epsilon_{a_1 \dots a_{d-2}} t^a v_a
\equiv -t^a v_a \sqrt{\sigma}d^{d-2}x \,
\end{equation}
gives
\begin{equation}
\label{hamgen}
\H_\xi = -\int_{S_\infty} \tilde{E}_{ab} t^a \xi^b \sqrt{\sigma} \, d^{d-2} x
- (\lambda_+ - \lambda_-)\int_{S_\infty} \left[W(\alpha) - \frac{\lambda_-}{d-1} \alpha \beta\right] t^a \xi_a \sqrt{\sigma} \, d^{d-2} x \,.
\end{equation}
Apart from the factor of $(\lambda_+ - \lambda_-)$, this verifies the form of $\H_\xi$ conjectured in \cite{hh} for the cases considered thus far.

From the definition~(\ref{gendef}), it is clear that $\H_\xi$ is defined only up to a term whose variation vanishes.  We have made a particular
choice of this term in (\ref{hamgen}).  Since the Weyl tensor vanishes in exact AdS space, our choice sets $\H_\xi = 0$ there when $W(0)=0$.

Let us consider the action of a conformal transformation on expression~(\ref{hamgen}).  A general choice of boundary conditions breaks the AdS symmetry
and is not invariant under rescaling of the conformal factor.  Boundary conditions preserving these symmetries may be found by considering the rescaling
$\Omega \to \omega \Omega$.  Since this changes only the auxiliary structure $\Omega$, the physical field $\phi$ must remain
invariant.
Thus, we must have $\alpha \to \omega^{-\lambda_-} \alpha$ and $\beta \to \omega^{-\lambda_+} \beta$.
This result and eq.~(\ref{dW}) then imply that the requirement for conformally invariant boundary conditions is
\begin{equation}
\label{winvt}
W(\alpha) = k \alpha^{(d-1)/\lambda_-} \,.
\end{equation}
Inserting this into eq.~(\ref{hamgen}), we find that the contributions from the scalar fields vanish.  The remaining term involving the Weyl tensor
is conformally invariant.  Thus, for such cases, $\H_\xi$ is invariant as well.

For the special cases $4\lambda_- \leq d-1$, the $(\tilde h_{ab})_{2\lambda_-}$ and $(\tilde h_{ab})_{4\lambda_-}$ terms will give
a contribution to $\delta \H_\xi$ at $O(\Omega^{4\lambda_- +1 -d})$ and proportional to $\alpha^4$.
For $4\lambda_- < d-1$, these contributions are divergent as $\Omega \to 0$, but once again it turns out that
all such divergences cancel in the end.  Therefore, the result~(\ref{hamgen}) still holds.
For $d=4$ and $4\lambda_- = 3$, the $\alpha^4$ terms do not completely cancel in general and the expression for $\H_\xi$ becomes
\begin{eqnarray}
\label{hamgen2}
\H_\xi = -\int_{S_\infty} \tilde{E}_{ab} t^a \xi^b \sqrt{\sigma} \, d^{2} x
- \frac{3}{2} \int_{S_\infty} &\Bigg[&W(\alpha) - \frac{1}{4} \alpha \beta
\nonumber \\ &&+ \left(\frac{3}{128}+ \frac{c_4}{18}\right) \alpha^4 \Bigg] t^a \xi_a \sqrt{\sigma} \, d^2 x \,.
\end{eqnarray}
Here the extra $\alpha^4$ contribution is finite.  Under the rescaling $\Omega \to \omega \Omega$, the transformations required for $\phi$ to remain invariant
are now $\alpha \to \alpha \omega^{-\lambda_-}$ and $\beta \to (\beta -\gamma_2 \alpha^3 \, \textrm{log} \, \omega)\omega^{-\lambda_+}$.
Conformally invariant boundary conditions are then given by
\begin{equation}
W(\alpha) = k\alpha^4 + \frac{\gamma_2}{4\lambda_-} \, \alpha^4 \, \textrm{log} \, \alpha \,.
\end{equation}
Inserting this into eq.~(\ref{hamgen2}) again cancels all contributions from the scalar fields, leaving only the Weyl curvature term.
Despite the apparent disparity between eqs.~(\ref{hamgen2})
and (\ref{hamgen}), it can be shown that~(\ref{hamgen2}) is actually the limit of~(\ref{hamgen}) as $4\lambda_- \to 3$.  This calculation is given
in appendix A.

In a similar way, we can also obtain an expression for the Hamiltonian generator in the case $m^2 = m^{2}_{BF}$ by considering
eq.~(\ref{hamgen}) in the limit $(\lambda_+ - \lambda_-) \equiv \epsilon\to 0$.
Eq.~(\ref{phi}) can be rewritten as
\begin{equation}
\phi = \Omega^{\lambda_+} (\alpha \Omega^{-\epsilon} + \beta) \,.
\end{equation}
Using $\Omega^{-\epsilon} = 1- \epsilon \, \textrm{log} \, \Omega + \ldots$ and comparing to eq.~(\ref{phibf}), we see that in the limit $\epsilon \to 0$ we have
\begin{equation}
\alpha \to \frac{\hat{\alpha}}{\epsilon}, \quad \beta \to \hat{\beta} - \frac{\hat{\alpha}}{\epsilon},
\quad \epsilon W \to \hat{W} - \frac{\hat{\alpha}^2}{2\epsilon} \, .
\end{equation}
This then gives
\begin{equation}
\label{hamgenBF}
\H_\xi = -\int_{S_\infty} \tilde{E}_{ab} t^a \xi^b \sqrt{\sigma} \, d^{d-2} x
- \int_{S_\infty} \left[\hat{W}(\hat{\alpha}) - \frac{1}{2} \hat{\alpha} \hat{\beta} - \frac{\hat{\alpha}^2}{2(d-1)} \right] t^a \xi_a \sqrt{\sigma} \, d^{d-2} x \,,
\end{equation}
generalizing the $\hat \alpha = 0$ result of \cite{HIM}.  In appendix B, we derive the result (\ref{hamgenBF}) by direct calculation.
Conformally invariant boundary conditions are now given by
\begin{equation}
\hat W(\hat\alpha) = k\hat\alpha^2 -\frac{1}{2\lambda} \, \hat\alpha^2 \, \textrm{log} \, \hat\alpha \,.
\end{equation}
When this is substituted into eq.~(\ref{hamgenBF}), the terms involving the scalar fields once again cancel, leaving only the Weyl term.

\section{The Spinor Charge}
\label{spinor}

In this section we relate the Hamiltonian generator~(\ref{hamgen}) to the spinor charge.
The spinor charge is shown to be manifestly positive, which implies a lower bound
on the corresponding energy.

Let $\psi$ be a spinor field such that asymptotically $-\bar{\psi}\gamma^a\psi = \xi^a$.
The spinor charge is then defined as
\begin{equation}
\label{charge}
Q_\xi = \int_{S_\infty} \ast {\bf B} \,,
\end{equation}
where the integrand is the Hodge dual of the Nester two-form \cite{n},
\begin{equation}
B_{cd} = \bar{\psi} \gamma_{cde} \widehat{\nabla}^e \psi + \textrm{h.c.}\, ,
\end{equation}
and $S_\infty$ is a cut of $\I$ that bounds a surface $\Sigma$.
The antisymmetrized curved space gamma matrices have been written as $\gamma^{a_1 a_2 \ldots a_n}=\gamma^{[a_1} \gamma^{a_2} \ldots \gamma^{a_n]}$.
We also define the covariant derivative
\begin{equation}
\widehat{\nabla}_a \psi = \nabla_a \psi + f(\phi) \gamma_a \psi \, ,
\end{equation}
where $f$ is a function of the scalar field $\phi$ to be fixed later.
First, following \cite{t}, we show that $Q_\xi \geq 0$.  Using Gauss's theorem we can rewrite the spinor charge as
\begin{equation}
Q_\xi = \int_{\Sigma} d \ast {\bf B} = \int dS_a \nabla_b B^{ab} \,.
\end{equation}
Our gamma matrix conventions are $\gamma_{(a}\gamma_{b)} = g_{ab}, (\gamma^0)^\dagger = -\gamma^0$, and $(\gamma^i)^\dagger = \gamma^i$.
We choose the 0-direction to be normal to $\Sigma$ and let $i,j,\ldots$ denote indices tangent to the surface.  Then, a few steps of algebra
and gamma matrix identities lead to
\begin{equation}
\label{divB}
\nabla_b B^{0b} = 2(\widehat{\nabla}_i \psi)^\dagger \widehat{\nabla}^i \psi + R^0 \, ,
\end{equation}
where
\begin{equation}
R^a = -\bar{\psi} \gamma^{bac} \widehat{\nabla}_b \widehat{\nabla}_c \psi + \textrm{h.c} \,.
\end{equation}
To obtain eq.~(\ref{divB}), we have imposed the Witten condition \cite{witten2}
\begin{equation}
\label{witten}
\gamma^i \widehat{\nabla}_i \psi = 0 \,.
\end{equation}
The first term in eq.~(\ref{divB}) is clearly nonnegative as written, while the second term can be written in the form $\lambda^\dagger \lambda$, where
\begin{equation}
\lambda = \frac{1}{\sqrt{2}}\left(\gamma^a \nabla_a \phi -2(d-2)\frac{df}{d\phi}\right) \psi \,,
\end{equation}
and the function $f(\phi)$ must satisfy
\begin{equation}
\label{vtof}
V(\phi) = -2(d-1)(d-2)f^2 +2(d-2)^2 \left(\frac{df}{d\phi}\right)^2 \, .
\end{equation}
That is, $V(\phi)$ arises from a superpotential $f(\phi)$.  Assuming a series solution for $f$ and using our expression for $V(\phi)$ gives
\begin{equation}
\label{spotential}
f(\phi) = \frac{1}{2} +\frac{\lambda_-}{4(d-2)} \phi^2 + \frac{\gamma_1 \lambda_-}{8} \phi^4 + \ldots \,.
\end{equation}
An asymptotic series of the form (\ref{spotential}) solves (\ref{vtof}) for any $V$.
However, positivity of $Q_\xi$ is assured only if $f(\phi)$ exists for all $\phi$.
We assume below that this is the case, and that this global solution is of the form (\ref{spotential}).  See \cite{new} for further discussion of this point.

To calculate the spinor charge explicitly we need to expand $\widehat{\nabla}_a \psi$ in powers of $\Omega$.  We write $\psi = \psi_0 + \ldots$,
where $\widehat{\nabla}_a \psi_0 = 0$ (i.e. $\psi_0$ is a Killing spinor) in exact AdS space.  Higher order terms in the
expansion are determined by solving eq.~(\ref{witten})
order by order (see the appendix of \cite{hh}).  We find
\begin{equation}
\label{psi}
\psi =\psi_0 + \psi_{2\lambda_-}\Omega^{2\lambda_-} + \ldots,
\end{equation}
where
\begin{equation}
\psi_{2\lambda_-} = -\frac{\alpha^2}{8(d-2)} \, \psi_0\,.
\end{equation}
Define $\tilde \psi = \Omega^{1/2} \psi$ and $\tilde \gamma_a = \Omega \gamma_a$.  Then, under a conformal transformation $\tg_{ab} = \Omega^2 g_{ab}$, one has
\begin{equation}
\nabla_a \psi = \tnabla_a \psi - \frac{1}{2} \Omega^{-1} (\tgam_a \tgam_b \tn^b - \tn_a) \psi \,,
\end{equation}
so
\begin{equation}
\label{delpsi}
\Omega^{1/2} \widehat{\nabla}_a \psi = (\tnabla_a)_0 \tilde{\psi} + \tilde{\Gamma}_a \tilde{\psi} + \frac{1}{2} \Omega^{-1}
\tgam_a(1-\tgam_b \tn^b) \tilde{\psi} - \frac{1}{2} \Omega^{-1} \tgam_a (1-2f(\phi)) \tilde{\psi} \,,
\end{equation}
where $(\tnabla_a)_0$ is the covariant derivative with respect to pure AdS space and $\tilde{\Gamma}_a$ is the linearized spin connection.
Then eqs.~(\ref{phi}), (\ref{ds}), (\ref{spotential}), and (\ref{psi}) give
\begin{eqnarray}
\label{delpsi2}
\Omega^{2-d}[\Omega^{1/2}\widehat{\nabla}_e \psi] &=& -\frac{\lambda_- \alpha^2}{4(d-2)} \Omega^{2\lambda_- + 1 -d} \tilde{h}_{ef} \tgam^f \tilde{\psi}_0
+\frac{\lambda_- \alpha^2}{4(d-2)} \Omega^{2\lambda_- + 1 -d} \tgam_e \tilde{\psi}_0 \nonumber\\ &&-
\frac{1}{2} \tilde{E}_{ef} \tgam^f \tilde{\psi}_0 +
\frac{\alpha \beta}{2(d-2)}\left(\frac{2m^2}{d-1} \tilde{h}_{ef} \tgam^f + \lambda_- \tgam_e \right) \tilde{\psi}_0 + \ldots,
\end{eqnarray}
where here the \ldots represents additional terms which will not contribute to the spinor charge.  A number of these terms are proportional to
$\tn_e$, and any such term vanishes when inserted into eq.~(\ref{charge}).  The first and second terms in eq.~(\ref{delpsi2})
arise from the second and last terms respectively in eq.~(\ref{delpsi}), and are divergent as $\Omega \to 0$.  When $4\lambda_- < 3$, there
are also divergent terms proportional to $\alpha^4 \Omega^{4\lambda_- -3}$, which we have not written down in the interest of brevity.
Next we insert the above expression into the definition of the spinor charge and use (on $\I$)
\begin{eqnarray}
\bar{\tilde{\psi}} \, \tilde{\Gamma}^{cde}  \tilde{E}_{ef} \tgam^f \tilde{\psi}_0 \tilde{\epsilon}_{cd a_1 \ldots a_{d-2}} &=&
-2 \tilde{E}^d{}_f \xi^f \tn^c \tilde{\epsilon}_{cd a_1 \ldots a_{d-2}} \\
\bar{\tilde{\psi}} \, \tilde{\Gamma}^{cde} \tgam_e \tilde{\psi}_0 \tilde{\epsilon}_{cd a_1 \ldots a_{d-2}} &=&
2(d-2) \xi^d \tn^c \tilde{\epsilon}_{cd a_1 \ldots a_{d-2}}
\end{eqnarray}
to arrive at
\begin{equation}
\label{qtoh}
Q_\xi = \H_\xi  + (\lambda_+ - \lambda_-)\int_{S_\infty} W(\alpha) \sqrt{\sigma} \, t^a \xi_a d^{d-2} x \,,
\end{equation}
where $\H_\xi$ is given in eq.~(\ref{hamgen}).  Once again, all the divergent terms have canceled in the final result.

The lower bound on the energy now follows immediately.
Let $\xi^a = t^a + \omega \varphi^a$, where $\varphi^a$ is a sum of vectors generating spatial rotations and $\omega$ is a real number such that $|\omega| < 1$.
Then $\H_\xi = E +\omega J$, where $E$ is the energy and $J$ is the angular momentum.  Use of $Q_\xi \geq 0$ and the relation~(\ref{qtoh})
then show that the energy is bounded below when $W$ has a global minimum, with the bound given by
\begin{equation}
\label{bound}
E \geq \frac{2 \pi^{(d-1)/2}}{\Gamma\left(\frac{d-1}{2}\right)} (\lambda_+ - \lambda_-) \, \textrm{inf} \, W + |J| \,.
\end{equation}

The case $d=4$ and $4\lambda_- = 3$ evidently requires special treatment, since the factor $\gamma_1$ in eq.~(\ref{spotential}) diverges in this limit.
For this case, we can choose instead the superpotential
\begin{equation}
\label{2f}
f(\phi) = \frac{1}{2} + \frac{3}{32} \phi^2 + \frac{\gamma_2}{8} \phi^4 \, \textrm{log} \, \phi + \ldots \,,
\end{equation}
which satisfies eq.~(\ref{vtof}) when $4\lambda_- =3$.  Again, we assume that a superpotential of this form exists for all $\phi$.
Note that although we write $V(\phi)$ in terms of the
``superpotential'' $f(\phi)$, we have not required our system to be
supersymmetric.  Due to the non-analyticity of the $\textrm{log} \, \phi$ term, one expects $\gamma_2 =0$ in sufficiently
supersymmetric theories.

Carrying out the same steps as above leads to
\begin{equation}
Q_\xi = \H_\xi  + \frac{3}{2} \int_{S_\infty} {\mathcal W}(\alpha) \sqrt{\sigma} \, t^a \xi_a d^{d-2} x \,,
\end{equation}
where $\H_\xi$ is given in eq.~(\ref{hamgen2}) and
\begin{equation}
\label{bound2}
{\mathcal W}(\alpha) = W(\alpha) + \left(\frac{3}{64}+\frac{c_4}{9}\right)(1 - 2 \, \textrm{log} \, \alpha) \alpha^4 \,.
\end{equation}
Positivity of the spinor charge then implies a lower bound on the energy of the form~(\ref{bound}) with $W \to {\mathcal W}$.

When $m^2 = m^{2}_{BF}$, the spinor charge is equal to $\H_\xi$ as given in eq.~(\ref{hamgenBF}) plus a divergent term.  This can
be shown by direct calculation, or by considering eq.~(\ref{qtoh}) in the limit $\epsilon = (\lambda_+-\lambda_- )\to 0$.  Thus, in this
case, we do not find a lower bound on $E$.

\section{Discussion}
\label{disc}

We have studied conserved charges in asymptotically anti-de Sitter spacetimes containing scalar fields with mass in the range~(\ref{range}).
Our first main result is that our choice of asymptotic conditions leads to Hamiltonian generators of asymptotic symmetries which are finite and well defined.
Our results agree with those of \cite{hmtz}, obtained by notably different methods.  We gave a general argument for finiteness in section~\ref{structure},
which was later confirmed by our derivation of explicit expressions for the generators under the assumption that the scalar potential $V(\phi)$ is even.  We note
that \cite{hmtz} allows odd terms in  $V(\phi)$, in addition to the even terms considered here.

For generic cases, our explicit form for the generators~(\ref{hamgen}) verifies the conjecture of \cite{hh} (up to a factor of $(\lambda_+ - \lambda_-)$),
which was based on the idea of
conformal invariance.  We did, however, obtain a different
form for $\H_\xi$ in two special cases where the scalar field solution has a logarithmic branch.
When $4\lambda_- = 3$, we find in eq.~(\ref{hamgen2}) that there is an additional contribution proportional to $\alpha^4$.
This is the marginal case at which the $\phi^4$ term in the potential becomes relevant, and we would expect similar corrections from $\phi^3$ or $\phi^5$
terms were they to be included.  The other interesting case occurs when the BF bound is saturated, with $\H_\xi$ now given by eq.~(\ref{hamgenBF}).
We have considered only minimally coupled scalar fields, but we would expect that the general argument in favor of finite generators
applies equally well in the non-minimal case.

The second key component of this work is use of the spinor charge to show that the energy is
bounded below when i) $W$ has a global minimum,  ii) $m^2 > m^2_{BF}$, and iii) $V(\phi)$ admits a superpotential of the form (\ref{spotential}) or (\ref{2f}).
See \cite{new,counter} for further discussion of requirement (iii).  In particular, as discussed in \cite{new}, not all superpotentials $f$ solving (\ref{vtof})
 are in fact of the form (\ref{spotential}), (\ref{2f}).

The detailed form of the bound is slightly modified in the case $4\lambda_- =3$ and
our method yields no such bound when $m^2 = m^2_{BF}$.   An expression for the bound when it exists is given in eq.~(\ref{bound}).
This would then imply that the theory has a stable ground state, as predicted by AdS/CFT.  Note however, that \cite{iw} found stability in the corresponding
linear theory even in some cases where $W$ is not bounded below.  This suggests that our results could be improved and that further studies of
stability in the dual CFT should also be performed.

\begin{acknowledgments}
We are pleased to thank Sean Hartnoll, Marc Henneaux, Stefan Hollands, Gary Horowitz, and Akihiro Ishibashi for valuable discussions.  We are especially grateful
to Marc Henneaux, Cristi\'an Mart\'inez, Ricardo Troncoso, and Jorge Zanelli for providing advance access to their results, which allowed us to correct an error in an
early draft.  This work was supported in part by NSF grants PHY0354978 and PHY99-07949, and by funds from the University of California.
D.M. would also like to thank the Kavli Institute
for Theoretical Physics for its hospitality during the final stages of this work.
\end{acknowledgments}

\appendix

\section{Hamiltonian Generator for $4\lambda_- =3$}
In this appendix, we show how to obtain (\ref{hamgen2}) from (\ref{hamgen}) in the limit
\begin{displaymath}
\varepsilon \equiv (4\lambda_- -3) \to 0 \,.
\end{displaymath}
As argued above, it is only necessary to consider $d=4$.  Recall that the expansion of the scalar field near infinity for  $4\lambda_- < 3$ is
\begin{equation}
\label{phi1}
\phi = \alpha \Omega^{\lambda_{-}} + \gamma_1 \alpha^3 \Omega^{3\lambda_{-}} + \beta \Omega^{\lambda_{+}}  +\ldots \,,
\end{equation}
while for $4\lambda_- = 3$ we have
\begin{equation}
\label{phi2}
\phi = \bar \alpha \Omega^{\lambda_{-}} + \gamma_2 \bar \alpha^3 \Omega^{3\lambda_{-}} \, \textrm{log} \, \Omega + \bar \beta \Omega^{\lambda_{+}}+ \ldots \,.
\end{equation}
Here we have been careful to distinguish the coefficients of the $\Omega^{\lambda_\pm}$ solutions in the two cases.
We then observe that eq.~(\ref{phi1}) can be written in the form
\begin{equation}
\phi = \alpha \Omega^{\lambda_{-}} +  \Omega^{\lambda_{+}}(\gamma_1 \alpha^3 \Omega^{\varepsilon} + \beta)  +\ldots \,.
\end{equation}
Using $\Omega^\varepsilon = 1 + \varepsilon \,\textrm{log} \, \Omega + \ldots$ and comparing to eq.~(\ref{phi2}), we see that in the limit $\varepsilon \to 0$
one obtains
\begin{equation}
\alpha \to \bar \alpha, \quad \beta \to \bar\beta - \frac{\gamma_2}{\varepsilon}\,  \bar \alpha^3, \quad \gamma_1 \to \frac{\gamma_2}{\varepsilon} \,.
\end{equation}
The consistency of the third limit can be checked from eqs.~(\ref{gam1}) and (\ref{gam2}).
The boundary condition~(\ref{dW}) then implies that
\begin{equation}
W \to \bar W -\frac{\gamma_2}{4\varepsilon} \, \bar \alpha^4 \,.
\end{equation}
In taking the $\varepsilon \to 0$ limit of eq.~(\ref{hamgen}), we write $4\lambda_- = 3+ \varepsilon$ and find that we must make the replacement
\begin{equation}
(\lambda_+-\lambda_-)\left(W - \frac{\lambda_-}{3} \alpha \beta\right) \to \frac{3}{2}\left( \bar W -\frac{1}{4} \bar\alpha \bar\beta +
\frac{\gamma_2}{12} \bar\alpha^4 \right) \,.
\end{equation}
Finally, using $\gamma_2 = 9/32 + 2c_4/3$ we obtain that for $4\lambda_- = 3$ the generator is
\begin{eqnarray}
\H_\xi = -\int_{S_\infty} \tilde{E}_{ab} t^a \xi^b \sqrt{\sigma} \, d^{2} x
- \frac{3}{2} \int_{S_\infty} &\Bigg[& \bar W(\bar \alpha) - \frac{1}{4} \bar \alpha \bar \beta
\nonumber \\ &&+ \left(\frac{3}{128}+ \frac{c_4}{18}\right) \bar \alpha^4 \Bigg] t^a \xi_a \sqrt{\sigma} \, d^2 x \,.
\end{eqnarray}
Renaming coefficients then produces eq.~(\ref{hamgen2}) as claimed.

\section{Hamiltonian Generator for $m^2 = m^2_{BF}$}
In this appendix, we give a more detailed calculation of the Hamiltonian generator when the BF bound is saturated; that is, when
\begin{equation}
m^2 = m^{2}_{BF} = -\frac{(d-1)^2}{4} \,.
\end{equation}
The asymptotic behavior of the scalar field in this case is
\begin{equation}
\phi = -\hat{\alpha} \Omega^{\lambda}\,\textrm{log}\,\Omega + \hat{\beta} \Omega^{\lambda} + \dots \, ,
\end{equation}
where $\lambda = (d-1)/2$.
Inserting this into the stress energy tensor, eq.~(\ref{Labdef}), gives
\begin{eqnarray}
\label{LabBF}
\tilde L_{ab} &=& \frac{2 \lambda^2 \hat{\alpha}^{2}}{d-2} \left[\tn_{a} \tn_{b} - \frac{1}{d-1} \tilde{g}_{ab}\right]
\Omega^{d-3}\,(\textrm{log}\,\Omega)^2 \cr
&& + \frac{2\lambda \hat\alpha}{d-2} \left[2(\hat\alpha - \lambda \hat\beta)\tn_{a} \tn_{b} - \frac{\hat\alpha - 2\lambda \hat\beta}{d-1} \tilde{g}_{ab}\right]
\Omega^{d-3}\,\textrm{log}\,\Omega \nonumber\\
&& +\left[ \frac{2(\hat\alpha - \lambda \hat\beta)^2}{d-2} \tn_{a} \tn_{b} -\frac{\hat\alpha^2-2\lambda \hat\alpha \hat\beta +2 \lambda^2\hat\beta^2}{(d-2)(d-1)}
\tilde{g}_{ab}\right] \Omega^{d-3} + \ldots \,.
\end{eqnarray}
This suggests that now the series expansions should be of the form
\begin{equation}
\label{seriesBF}
\tilde h_{ab} = \sum_{i,j}\Omega^{i} \, (\textrm{log} \, \Omega)^j (\tilde h_{ab})_{i,j}, \quad
\tilde K = \sum_{i,j} \Omega^{i} \, (\textrm{log} \, \Omega)^j (\tilde K)_{i,j}, \ldots \textrm{etc.},
\end{equation}
where the indices $i,j$ run over nonnegative integers.  The recursion relations are similar to those given above, but there are extra terms due to the
logarithms:
\begin{eqnarray}
\label{recursionpBF}
(i-d+2)(\tilde p_a{}^b)_{i,j} &=& -(j+1)(\tilde p_a{}^b)_{i,j+1}
-(\tilde \R_a{}^b)_{i-1,j} + \frac{1}{d-1}(\tilde \R)_{i-1,j} \delta_a{}^b
\nonumber \\
&&+(\tilde \K \tilde p_a{}^b)_{i-1,j} +\frac{d-2}{2} (\tilde \tau_a{}^b)_{i-1,j}
\\
\label{recursionkBF}
(i-2d+3)(\tilde \K)_{i,j} &=& -(j+1)(\tilde \K)_{i,j+1} -(\tilde \R)_{i-1,j}
+ (\tilde \K^2)_{i-1,j}
\nonumber \\
&&+\frac{d-1}{2} (\tilde L_{cd})_{i-1,j} \tn^c \tn^d + \frac{2d-3}{2}  (\tilde{L}_{cd} \tilde{h}^{cd})_{i-1,j}
\end{eqnarray}
\begin{equation}
\label{recursionhBF}
i(\tilde h_{ab})_{i,j}
= - (j+1)(\tilde h_{ab})_{i,j+1} -2(\tilde h_{bc} \tilde p_a{}^c)_{i-1,j}
         - \frac{2}{d-1} (\tilde \K \tilde h_{ab})_{i-1,j}.
\end{equation}
Then, eqs.~(\ref{LabBF}) and (\ref{recursionkBF}) indicate that the lowest order at which the nonzero stress energy tensor affects $\tilde K$
is $(\tilde{K})_{d-2,2}$.  The $\tilde \R$ and $\tilde{K}^2$ terms must vanish because they are the same as in the $T_{ab} = 0$ case,
so one finds
\begin{equation}
(\tilde{K})_{d-2,2} = \frac{\lambda^2}{d-2} \hat\alpha^2
\end{equation}
and by analogous arguments
\begin{equation}
(\tilde{K})_{d-2,1} = \frac{2\lambda^2}{d-2} \left(\frac{\hat\alpha^2}{d-1}-\hat\alpha \hat\beta\right) \,.
\end{equation}
Eq.~(\ref{recursionpBF}) implies $(\tilde p_a{}^b)_{i,j}$ vanishes at these orders so eq.~(\ref{recursionhBF}) gives
\begin{equation}
(\tilde h_{ab})_{d-1,2} = -\frac{\hat\alpha^2}{2(d-2)} (\tilde h_{ab})_{0,0} \quad \textrm{and} \quad
(\tilde h_{ab})_{d-1,1} = \frac{\hat\alpha \hat\beta}{d-2} (\tilde h_{ab})_{0,0} \,.
\end{equation}
Once again, we write $(\tilde h_{ab})_{d-1,0}$ in terms of the electric part of the Weyl tensor.  Expansions of the form~(\ref{seriesBF}) modify eq.~(\ref{habd}) to be
\begin{eqnarray}
(\tilde{h}_{ab})_{d-1,0} &=& -\frac{2}{d-1} (\tilde{E}_{ab})_{0,0} + \frac{2}{(d-1)(d-3)} (\tilde{K}_{ac} \tilde{K}^c{}_b)_{d-3,0}
\nonumber\\ &&- \frac{1}{(d-1)(d-3)}(\tilde h_a{}^c \tilde h_b{}^d \tilde L_{cd} + \tilde h_{ab} \tilde L_{cd} \tn^c \tn^d)_{d-3,0}
-\frac{2}{(d-1)(d-3)} (\tilde h_{ab})_{d-1,2} \nonumber\\
&&-\frac{2(d-2)}{(d-1)(d-3)}(\tilde h_{ab})_{d-1,1} \,.
\end{eqnarray}
The expression for the metric near infinity is then
\begin{eqnarray}
\label{dsBF}
d\tilde s^2 &=& d\Omega^2 -\Bigg[\left(1 + \frac{1}{4} \Omega^2 \right)^2 - \frac{\hat\alpha^2}{2(d-2)}
\Omega^{d-1}\,(\textrm{log}\, \Omega)^2 + \frac{\hat\alpha \hat\beta}{d-2}
\Omega^{d-1} \,\textrm{log}\, \Omega \nonumber \\
&&-\frac{1}{2(d-2)} \left(\frac{2\hat\alpha^2}{(d-1)^2}+\hat\beta^2\right) \Omega^{d-1} \Bigg] dt^2
+\Bigg[\left(1 - \frac{1}{4} \Omega^2 \right)^2 - \frac{\hat\alpha^2}{2(d-2)}
\Omega^{d-1}\,(\textrm{log}\, \Omega)^2 \nonumber \\
&&+ \frac{\hat\alpha \hat\beta}{d-2}
\Omega^{d-1} \,\textrm{log}\, \Omega -\frac{1}{2(d-2)} \left(\frac{2\hat\alpha^2}{(d-1)^2}+\hat\beta^2\right)\Omega^{d-1}\Bigg]  d\sigma^2
\nonumber\\ &&- \frac{2}{d-1}\Omega^{d-1} \tilde{E}_{ab} \, dx^a dx^b + \dots \,.
\end{eqnarray}

Once again the consistency condition for the existence of the Hamiltonian generator is satisfied since the pullback of $\boldsymbol{\omega}$ to $\I$ vanishes.
Now consider the metric variation
\begin{eqnarray}
\delta \tilde{g}_{ab} &=& - \frac{\delta(\hat\alpha^2)}{2(d-2)}
\Omega^{d-1}\,(\textrm{log}\, \Omega)^2 (\tilde h_{ab})_0
+ \frac{\delta(\hat\alpha \hat\beta)}{d-2}
\Omega^{d-1} \,\textrm{log}\, \Omega (\tilde h_{ab})_0
\nonumber\\ && -\frac{1}{2(d-2)} \left(\frac{2\delta(\hat\alpha^2)}{(d-1)^2}+\delta(\hat\beta^2)\right) \Omega^{d-1} (\tilde h_{ab})_0
 -\frac{2}{d-1} \Omega^{d-1} \delta \tilde{E}_{ab} \,.
\end{eqnarray}
In general, a metric variation of the form
\begin{displaymath}
\delta \tg_{ab} = b(\delta \hat \alpha, \delta \hat \beta) \,\Omega^{d-1} \, (\textrm{log} \,\Omega)^n (\tilde h_{ab})_0
\end{displaymath}
yields
\begin{displaymath}
(\delta Q_\xi -\xi \cdot \theta^G)_{a_1 \dots a_{d-2}} \mid_{\I} \, = \frac{(d-2)b}{2} \, [(d-1)(\textrm{log} \,\Omega)^n + n (\textrm{log} \,\Omega)^{n-1}]\tn^b \xi^c
\tilde \epsilon_{bca_1 \dots a_{d-2}} \,,
\end{displaymath}
where $b$ is some function of $\delta \hat \alpha$ and $\delta \hat \beta$.
For $n \neq 0$, the above result has terms diverging as $\Omega \to 0$.  These turn out to be canceled by similar divergences arising in the expression for
$\xi \cdot \boldsymbol{\theta}^\phi$, given by
\begin{equation}
(\xi \cdot \theta^\phi)_{a_1 \dots a_{d-2}} = -\left[\frac{\lambda}{2} \delta(\hat \alpha^2) (\textrm{log} \,\Omega)^2 +
\left( \frac{\delta(\hat \alpha^2)}{2} - \lambda \delta(\hat \alpha \hat \beta)\right)
\textrm{log} \,\Omega -\hat \alpha \delta \hat\beta + \frac{\lambda}{2} \delta(\hat\beta^2)\right]    \tn^b \xi^c
\tilde \epsilon_{bca_1 \dots a_{d-2}} \,.
\end{equation}
The final result for the Hamiltonian generator is then
\begin{equation}
\H_\xi = -\int_{S_\infty} \tilde{E}_{ab} t^a \xi^b \sqrt{\sigma} \, d^{d-2} x
- \int_{S_\infty} \left[\hat{W}(\hat{\alpha}) - \frac{1}{2} \hat{\alpha} \hat{\beta} - \frac{\hat{\alpha}^2}{2(d-1)} \right] t^a \xi_a \sqrt{\sigma} \, d^{d-2} x \,,
\end{equation}
which verifies eq.~(\ref{hamgenBF}).

\end{document}